\documentclass[twocolumn,aps,pre,
groupedaddress,amssymb,amsmath,nobalancelastpage]{revtex4}

\usepackage{graphicx}% Include figure files
\usepackage{longtable}% Include figure files

\begin{document}

\title{Velocity Fluctuations in a Slowly Sheared Bubble Raft}

\author{Michael Dennin}
\affiliation{Department of Physics and Astronomy, University of
California at Irvine, Irvine, California 92697-4575}

\date{\today}

\begin{abstract}

A surprising feature of flow in slowly sheared model foam (bubble
raft) is a measured discontinuity in the rate of strain as a
function of position such that part of the system is ``flowing''
and the rest is undergoing ``elastic'' deformations [J. Lauridsen,
G. Chanan, and M. Dennin, Phys. Rev. Lett. {\bf 93}, 018303
(2004)]. Detailed measurements of the distribution of nonlinear
bubble rearrangements have been reported in connection with this
discontinuity. In this paper, measurements of the fluctuations in
velocity under the same conditions are reported. The fluctuations
are characterized by the second and third moments of the velocity
distribution. A surprising feature is the qualitative behavior of
these moments as a function of position in the system, especially
across the discontinuity in rate of strain. In addition, the
measured dependence of the second moment of the velocity
fluctuations on rate of strain is compared with predictions of
simulations of the bubble model and reasonable agreement is found.

\end{abstract}

\maketitle

%\begin{keyword}
%foam \sep velocity fluctuations \sep bubble raft \sep shear
%localization \sep complex fluids
%\end{keyword}
%\end{frontmatter}

\section{Introduction}

Bubble rafts (single layers of bubbles floating on a water
surface) serve as a model two-dimensional foam \cite{BL49,AK79}.
When subjected to sufficiently slow rates of strain, a bubble raft
exhibits an interesting discontinuity in the rate of strain
($\dot{\gamma}$) as a function of position \cite{LCD04}. Similar
discontinuities in $\dot{\gamma}$ have been observed for high
rates of strain in a number of other complex fluids
\cite{CRBMGH02,DCBC02}. This is in contrast to cases of flow
localization in which the rate of strain is continuous, as has
been observed in granular systems \cite{HBV99,MDKENJ00,LBLG00} and
confined two-dimensional foams \cite{DTM01}. The discontinuity of
strain rate in bubble rafts was observed for flow between two
concentric cylinders (Couette flow). At a critical radius ($r_c$),
the system is divided into a ``flowing'' region for $r < r_c$ and
a region undergoing ``elastic'' deformations for $r > r_c$.
Because of the concentric cylinder geometry, the immediate
question raised by the measurements in Ref.~\cite{LCD04} is the
connection between the discontinuity in $\dot{\gamma}$ and the
yield stress of the bubble raft. For both general foams and the
bubble raft, the yield stress is the critical stress below which
flow does not occur and the system acts as an elastic solid
\cite{K88,WH99}. Simple arguments suggest that the existence of
coexistence between flow and elastic behavior is not surprising
under the geometry used in Ref.~\cite{LCD04}. However, the same
arguments suggest that the rate of strain should be continuous
across the transition. The arguments depend on the combination of
a yield stress for the fluid and a spatially varying stress field
due to the Couette geometry \cite{BAH77}.

In the Couette geometry, the stress decreases as a function of
radial distance away from the inner cylinder \cite{BAH77}.
Therefore, for a yield stress fluid sheared in a Couette geometry,
one expects a region of flow close to the inner cylinder and, if
the stress drops below the yield stress, a critical radius above
which the system does not flow. Even for a yield stress fluid, it
is generally expected that the stress is a {\it continuous}
function of the rate of strain (e.g. Bingham plastic or
Herschel-Bulkley models of yield stress fluids \cite{BAH77}).
Therefore, one expects the {\it rate of strain} to be continuous
across the transition from flow to no flow. This is what makes the
observation of a discontinuity in rate of strain as a function of
position so surprising.

The discontinuity in the rate of strain was determined from
measurements of the average velocity of the bubbles as a function
of radial position \cite{LCD04}. It is interesting to ask how the
fluctuations in velocity behave near this discontinuity, and
whether or not the radial dependence of the velocity fluctuations
can shed any light on the source of the discontinuity. The
velocity fluctuations are interesting for another reason. In
granular systems, one often defines a ``granular temperature'' in
terms of the root mean squared value of the velocity. In this
context, there have been a number of studies of the dependence of
velocity fluctuations in a variety of granular flows
\cite{MD97,LCDKG99,LD00,M03}. It is interesting to ask if similar
ideas carry over to foam, and such a study has been done using
simulations of the bubble model for foam \cite{OTLL03}.

The bubble model is a simplified description of foam that has been
successful in capturing a number of observed features in bubble
rafts and other foam systems \cite{D95,D97,TSDKLL99}. The bubble
model treats foam as a collection of spherical bubbles. There are
only two forces in the bubble model that act on the bubbles. The
distortion of bubbles upon contact is accounted for by a repulsive
spring force proportional to the degree of overlap. Dissipation
within the foam is accounted for by including a viscous force
proportional to the velocity difference between two bubbles. These
two forces are balanced to determine the dynamics of the
individual bubbles. Using the bubble model, a detailed study of
velocity fluctuations as a function of packing fraction and
applied rate of strain was reported on in Ref.~\cite{OTLL03}. One
main result is the dependence of the second moment of velocity
fluctuations on rate of strain. It was found that the second
moment followed a power law as a function of rate of strain for
all values of rate of strain studied. This contrasted with a range
of alternate definitions of effective temperature, for which a
constant value of effective temperature was reached for
sufficiently low rate of strain \cite{OTLL03,OODLLN02}. Also of
interest to the work presented here is the fact that non-Gaussian
tails in the velocity distribution were measured for sufficiently
slow rates of strain.

In this paper, the velocity fluctuations for a bubble raft under
conditions of constant rate of strain in a Couette geometry are
reported. As discussed, under these conditions, there are two
regimes. Initially, the system acts as an elastic material. Above
a critical strain value, plastic flow occurs. The plastic flow is
characterized by irregular periods of stress increase and
decrease, with a constant average value of the stress
\cite{K88,WH99,LTD02}. It is during plastic flow that a
discontinuity in rate of strain as a function of radial position
in the system was observed \cite{LCD04}. In this paper, the second
and third moments of the velocity distributions are considered. A
number of results are discussed. First, for low rates of strain,
the second moment of the velocity distribution monotonically
decreases with decreasing rate of strain. Second, during flow,
there is a clear asymmetry in the velocity distribution that is
apparent in measurements of the third moment. Finally, comparison
is made between the velocity fluctuations during the initial
elastic deformation and in the zero rate of strain region during
flow.

\section{Experimental Methods}
\label{ExpMeth}

The experimental system has been previously described in some
detail \cite{app}. It consisted of a standard bubble raft
\cite{BL49,AK79} in a Couette geometry (two concentric cylinders
with a fluid confined in the region between the cylinders). For
the experiments discussed here, the outer radius was fixed at $r_o
= 7.43\ {\rm cm}$. The inner barrier, or rotor, is a Teflon disk
with a radius $r_i = 3.84\ {\rm cm}$. The outer edge of the disk
is a knife edge that is just in contact with the water surface. It
was suspended by a wire to form a torsion pendulum. The bubble
raft was produced by flowing regulated nitrogen gas through a
hypodermic needle into a homogeneous solution of of 80\% by volume
deionized water, 15\% by volume glycerine, and 5.0\% by volume
Miracle Bubbles (Imperial Toy Corp.). The bubble size was
dependent on the nitrogen flow rate, which was varied using a
needle valve.  A random distribution of bubble sizes was used,
with an average radius of $1\ {\rm mm}$. The resulting bubbles
were spooned into a cylindrical Couette viscometer. This produced
a two-dimensional model of a wet foam on a homogeneous liquid
substrate.

An important feature of the bubble raft is the gas area fraction.
To achieve a desired gas area fraction, the bubble raft was
constructed by placing the approximate number of desired bubbles
in the trough with the outer barrier set to a large radius. It is
important to note that the bubbles exhibited a strong attraction
to each other. The outer barrier was compressed until the desired
radius was reached. The gas area fraction was determined by
thresholding images of the bubbles and counting the area inside of
the bubbles. Because of the three-dimensional nature of the
bubbles, this represents an operational definition of gas-area
fraction based on the details of the image analysis. However, it
is also consistent with an estimate of the gas area fraction based
on the area of trough and expected distribution of bubble sizes.
For all of the data reported here, the gas area fraction was
approximately 0.95.

Flow is generated in the bubble raft by rotating the outer Teflon
barrier at a constant angular velocity. Results for two angular
velocities are reported: $\Omega = 8 \times 10^{-4}\ {\rm rad/s}$
and $\Omega = 5 \times 10^{-3}\ {\rm rad/s}$. The first layer of
bubbles at either boundary did not slip relative to the boundary.
Due to the finite size of the bubbles, this results in an
effective inner radius of $r = 4.4\ {\rm cm}$. Due to the
cylindrical geometry, the rate of strain is not uniform across the
system and is given by $\dot{\gamma} = r
\frac{d}{dr}\frac{v(r)}{r}$. Here $v(r)$ is the azimuthal velocity
of the bubbles. This allows for studies of the velocity
fluctuations over a wide range of rates of strain, even though
only two different rotation rates were used.

The details of the velocity measurements are given in
Ref~\cite{LCD04}. Video images of roughly one third of the trough
were recorded and individual bubbles motions were tracked. The
system is divided into equally spaced radially bins. Two different
types of radial bins are used. First, to compute the rate of
strain, bins of equal radial spacing are used. Second, to compute
properties as a function of the rate of strain, bins of equal
strain rate are used. Within each bin, the bubble tracks are used
to compute the average velocity, second moment of the velocity
distribution $<(v-\bar{v})^2>$, and the third moment of the
velocity distribution $<(v-\bar{v})^3>$ for the bin of interest.
In these expressions, the braces refer to an average over all
bubbles, and $\bar{v}$ is the mean velocity for the bin of
interest. For purposes of comparing with simulations, the variance
($\delta v = \sqrt{<(v-\bar{v})^2>}$) is considered, as well.

Before presenting the new results on the fluctuations, it is
useful to repeat what is already known from Ref.~\cite{LCD04}.
First, for $\Omega = 8 \times 10^{-4}\ {\rm rad/s}$, the critical
radius at which the rate of strain discontinuity occurs is $r_c =
6.7\ {\rm cm}$. For $\Omega = 5 \times 10^{-3}\ {\rm rad/s}$, $r_c
= 6.3 \ {\rm cm}$. For this paper, the region $r< r_c$ will be
referred to as the {\it flowing region}, and the region with $r >
r_c$ will be referred to as the {\it elastic deformation region}.
This is distinct from the difference between the {\it initial
elastic response} and the {\it steady-state flow}, which refer to
distinct time periods under steady strain. The rates of strains
used in this paper are taken from the same fits reported in
Ref.~\cite{LCD04}. The velocity data is fit in the region where
flow occurs, assuming a power law dependence of the velocity. This
fit is then used to determine the rate of strain by analytically
taking the derivative.

\section{Experimental Results}

\begin{figure}[h]
\includegraphics[width=8cm]{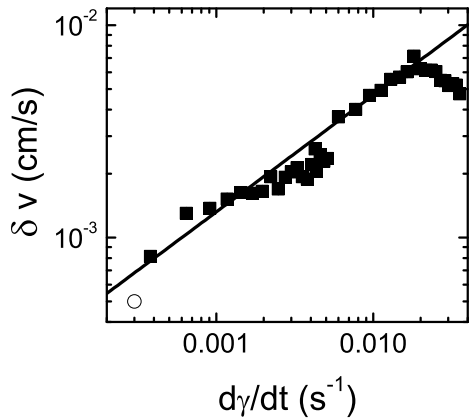}
\caption{\label{variance} A log-log plot of the variance ($\delta
v = \sqrt{<(v-\bar{v})^2>}$) as a function of rate of strain. The
squares are data taken from the flowing regime, and the open
circle is data taken from the initial elastic regime. The solid
line is the curve $\delta v = 0.59x^{0.55}$. At a rate of strain
of $0.02\ {\rm s^{-1}}$, there is a maximum in the variance. For
higher rates of strain, the variance decreases.}
\end{figure}

Figure~\ref{variance} shows the results for $\delta v =
\sqrt{<(v-\bar{v})^2>}$ as a function of shear rate. Both data
from $\Omega = 8 \times 10^{-4}\ {\rm rad/s}$ and $\Omega = 5
\times 10^{-3}\ {\rm rad/s}$ are shown. As a guide to the eye, the
solid line represents the curve $\delta v = 0.059x^{0.55}$. A
number of features of the behavior of $\delta v$ are worth noting.

First, the general trend for low rates of strain is consistent
with the simulations of the bubble model \cite{OTLL03}. Even the
exponent for the power law behavior is in reasonable agreement.
Assuming $\delta v \propto x^n$, the experimental data is
consistent with $n = 0.55 \pm 0.05$. The simulations give $n =
0.6$. The surprising feature is the decrease in $\delta v$ for
rates of strain above approximately $0.02\ {\rm s^{-1}}$. This is
a feature that is not reported in regard to the simulations.

\vspace{0.35in}

\begin{figure}
\includegraphics[width=8cm]{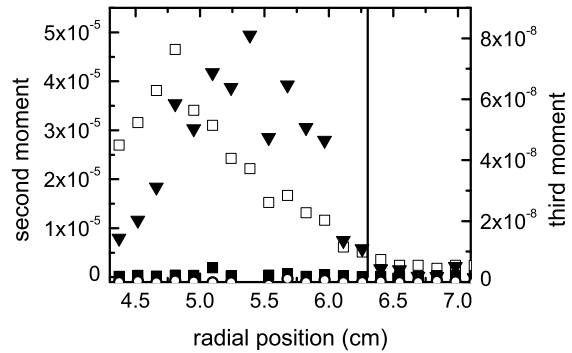}
\caption{\label{moments1} A plot of the second moment
($<(v-\bar{v})^2>$) of the velocity distribution versus radial
position in the system (left hand axis), and a plot of third
moment ($<(v-\bar{v})^3>$) of the velocity distribution versus
radial position in the system (right hand axis). All of the data
is for $\Omega = 5 \times 10^{-3}\ {\rm rad/s}$. The open squares
(second moment) and down triangles (third moment) are during flow,
and the solid squares (second moment) and open circles (third
moment) are during the initial elastic deformation. The vertical
line indicates the radial position of the rate of strain
discontinuity.}
\end{figure}

\begin{figure}
\includegraphics[width=8cm]{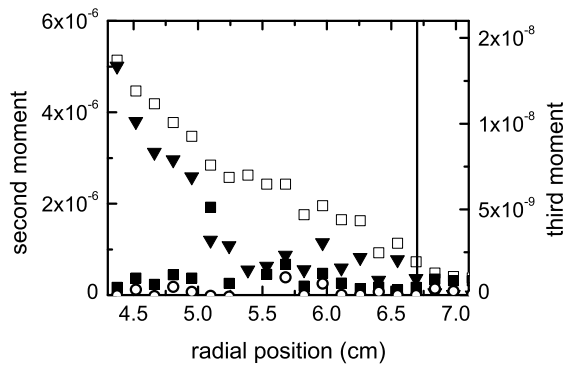}
\caption{\label{moments2} A plot of the second moment
($<(v-\bar{v})^2>$) of the velocity distribution versus radial
position in the system (left hand axis), and a plot of third
moment ($<(v-\bar{v})^3>$) of the velocity distribution versus
radial position in the system (right hand axis). All of the data
is for $\Omega = 8 \times 10^{-4}\ {\rm rad/s}$. The open squares
(second moment) and down triangles (third moment) are during flow,
and the solid squares (second moment) and open circles (third
moment) are during the initial elastic deformation. The vertical
line indicates the radial position of the rate of strain
discontinuity.}
\end{figure}

Figures~\ref{moments1} and \ref{moments2} illustrate the general
behavior of the second and third moment of the system as a
function of radial position for the two rotation rates studied:
$\Omega = 5 \times 10^{-3}\ {\rm rad/s}$ and $\Omega = 8 \times
10^{-4}\ {\rm rad/s}$, respectively. For comparison, the data for
the initial elastic deformation is included in each plot. This
data sets a baseline expectation for the measurements of
fluctuations. The main source of fluctuations during the flow is
expected to be nonlinear, neighbor switching events known as T1
events. These are rearrangements of bubbles in which bubbles
exchange neighbors. During the initial elastic flow, there are no
observed T1 events \cite{D04}. However, because of the method used
to measure stress, there is motion of the inner barrier. This
results in a small, non-zero rate of strain near the inner
boundary during the initial elastic flow \cite{D04}. Various
factors besides the T1 events contribute to velocity fluctuations
during this period, such as effects due to the finite size of the
bubbles and experimental noise.  For comparison to the
fluctuations during flow, the open circle in Fig.~\ref{variance}
is the value for $\delta v$ during this initial elastic period.

A number of features of the fluctuations are evident from the
plots in Figs.~\ref{moments1} and \ref{moments2}. First, recalling
that $\dot{\gamma}$ decreases as the radius increases for Couette
flow, the  $\dot{\gamma}$ dependence of the second moment is
apparent in Figs.~\ref{moments1} and \ref{moments2}. Second, there
is no obvious discontinuity in the second moment at $r_c$. For
$\Omega = 5 \times 10^{-3}\ {\rm rad/s}$, even in the elastic
deformation region, the second moment during flow is always
significantly larger than the second moment during the initial
elastic response. Finally, there is a dramatic increase in the
third moment in the flowing region. This indicates the development
of an asymmetry in the velocity distribution.

\begin{figure}
\includegraphics[width=8cm]{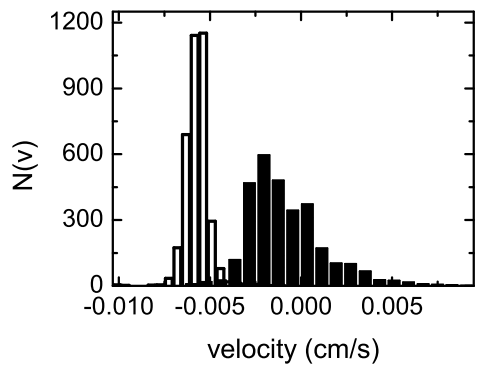}
\caption{\label{distr1} Distributions of velocities for $\Omega =
8 \times 10^{-4}\ {\rm rad/s}$. The two distributions are taken at
different radial positions. The open bars are for $r = 6.98\ {\rm
cm}$, and solid bars are for $r = 4.52\ {\rm cm}$. The two
distributions were selected to illustrate the evolution of an
asymmetry in the velocity distributions during flow.}
\end{figure}

The development of an asymmetry in the velocity distribution is
illustrated in Fig.~\ref{distr1}. Here the velocity distribution
is plotted for two different radial positions with $\Omega = 8
\times 10^{-4}\ {\rm rad/s}$. The solid bars are for $r = 4.52\
{\rm cm}$, and the open bars are for $r = 6.98\ {\rm cm}$. In
fact, one can see both the shift in average velocity and change in
the second moment, as well as the development of an asymmetric
distribution, for increasing $\dot{\gamma}$ (decreasing $r$).

\begin{figure}
\includegraphics[width=8cm]{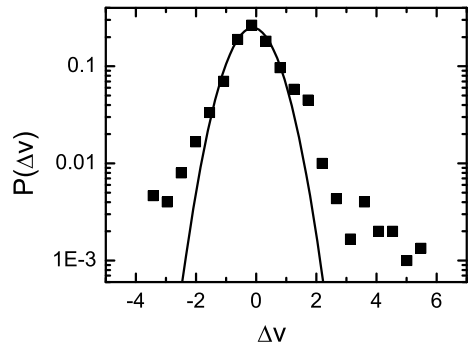}
\caption{\label{distr2} Comparison of the probability distribution
of $\Delta v = (v - \bar{v})/\delta v$ for a rate of strain of
$1.1 \times 10^{-3}\ {\rm s^{-1}}$ (solid squares) with a Gaussian
distribution (solid line). The data was taken at $r = 5.8\ {\rm
cm}$ with $\Omega = 8 \times 10^{-4}\ {\rm rad/s}$. The semi-log
plot highlights the non-Gaussian tails in the experimental
distribution.}
\end{figure}

Another feature of the distribution that is worth mentioning is
the observation of non-Gaussian tails. This is illustrated in
Fig.~\ref{distr2}. Here a typical probability distribution for the
difference in azimuthal velocity from the mean is plotted. The
case shown here is for a rate of strain of $1.1 \times 10^{-3}\
{\rm s^{-1}}$. For comparison with Ref.~\cite{OTLL03}, the data is
normalized by the variance, $\delta v$, so the probability
distribution for $\Delta v = (v - \bar{v})/\delta v$ is
considered. Also for comparison, a Gaussian distribution is
included in the plot. The use of a semi-log scale highlights the
non-Gaussian tails of the distribution.

\section{Discussion}

The bubble model has been successful at explaining many features
of slowly sheared bubble rafts, especially in relation to the
fluctuations in stress as the bubble raft flows \cite{LTD02,PD03}.
One feature of the bubble raft flow that is not immediately
obvious from the bubble model simulations is the existence of the
discontinuity in rate of strain \cite{LCD04}. The measurements of
velocity fluctuations presented here further explore this
discontinuity and various predictions of the bubble model.

There is basic agreement between the experimental measurements of
the second moment of the velocity fluctuations and the bubble
model. In both cases, the root mean squared fluctuations are
consistent with a power-law dependence on the rate of strain for
low rates of strain with an exponent of approximately 0.6
\cite{OTLL03}. The quantitative agreement is intriguing because
there was no attempt to match detailed features of the system to
the model, such as the dissipation or packing fraction. This
agreement raises the question of how robust this exponent is, and
what features of the system determine its value. Even further,
this exponent is in general agreement with similar measurements in
granular flow \cite{MD97,LCDKG99,LD00,M03}, providing further
support for a potential generalized description of these systems
in terms of a jamming transition \cite{LN98,ITPJamming}

A clear disagreement between the bubble model and the experiments
is the behavior of the second moment at high rates of strain. For
the experiments, the second moment of the velocity distribution
appears to decrease above a rate of strain of approximately $0.02\
{\rm s^{-1}}$. In contrast, the simulations observe a monotonic
increase in the second moment as a function of increasing rate of
strain to rates of strain of order 1 \cite{OTLL03}.

The sudden decrease in the second moment with rate of strain
raises two issues. First, it should be noted that in the
experiments, the decrease occurs at approximately the same rate of
strain for which it has been reported that the system makes the
transition to quasi-static behavior where the average stress is
essentially independent of rate of strain. In Ref.~\cite{PD03},
this was reported to occur at a rate of strain of $0.07\ {\rm
s^{-1}}$. However, the transition to quasi-static flow is not a
sharp transition, and a gap in the data in that work from $0.01\
{\rm s^{-1}} < \dot{\gamma} < 0.08\ {\rm s^{-1}}$ makes
determining its location difficult. Second, from
Fig.~\ref{moments1}, it is clear that the region of large rate of
strain is also a region relatively close to one of the system's
boundaries. This raises the issue of the impact of the boundaries
on fluctuations. Further work with larger systems will be needed
to resolve the role of boundaries.

The other clear agreement between the bubble model and the
experiments is the existence of non-Gaussian tails in the velocity
distribution \cite{OTLL03}. However, a result that needs to be
explored further is the asymmetry of the velocity distribution.
This asymmetry has not been reported for simulations of the bubble
model. However, the simulations did focus on the equivalent of the
radial velocity. In the experiments, the radial velocity was too
small for accurate measurements of the velocity distribution.
Therefore, before definite conclusions can be drawn, it is
necessary to consider the distribution for the velocity in the
direction of flow within the context of the bubble model. Besides
differences between velocities parallel and perpendicular to the
flow, the experiment and simulations use different geometries. It
may be that the asymmetry is a result of the Couette geometry, and
it does not exist in a parallel plate geometry. Future experiments
are planned for the parallel plate geometry that will help resolve
this issue.

Finally, the behavior near the discontinuity in strain rate raises
some interesting issues. Measurements of the average rate of
strain suggest that for $r > r_c$, the system is acting as an
elastic solid and that there is a discontinuity in the rate of
strain at $r_c$. However, measurements of the velocity
fluctuations clearly indicate a difference between the deformation
for $r > r_c$ during flow and the ``real'' elastic deformation
observed during the initial stages of rotation. First, there is no
obvious discontinuity in either the second or third moment of the
velocity distribution. Second, especially for $\Omega = 5 \times
10^{-3}\ {\rm rad/s}$, the second and third moments of the
velocity distribution, though essentially constant for $r > r_c$,
are significantly larger than the values observed during the
initial elastic rise. This indicates significant non-elastic
motion of the bubbles in this region. These results are consistent
with previous observations of nonlinear rearrangements, or T1
events. In Ref.~\cite{D04}, it was found that the T1 events occur
at all radii, even though the flow is localized to $r < r_c$. This
confirms the non-elastic nature of bubble motions during flow in
the region $r > r_c$. Again, future work on larger systems will be
needed to further increase our understanding of this interesting
discontinuity and to better understand the nature of the zero rate
of strain state for $r > r_c$ during flow.

\section{acknowledgments}

This work was supported by the Department of Energy grant
DE-FG02-03ED46071, the Research Corporation, and the Alfred P.
Sloan Foundation. I thank John Lauridsen for use of his video data
of bubble rafts.

\end{document}